# Anomalous Hall effect in Fe/Cu bilayers

W.J. Xu[1], B. Zhang[1], Z. Wang[1], S. Chu[1], W. Li[1], R.H. Yu[2] and X.X. Zhang[1]*

[1]*Dept. Phys. and Institute of Nanoscience & Technology, The Hong Kong University of Science and Technology, Clear Water Bay, Kowloon, Hong Kong, P. R. China.*

[2]*Laboratory of Advanced Materials, Department of Materials Science and Engineering, Tsinghua University, Beijing 100084, P. R. China*

**Abstract**

The scaling of anomalous Hall resistivity on the longitudinal resistivity has been intensively studied in the different magnetic systems, including multilayers and granular films, to examine which mechanism, skew scattering or side-jump, dominates. The basis of the scaling law is that both the resistivities are due to the electron scattering at the imperfections in the materials. By studying of anomalous Hall effect (AHE) in the simple Fe/Cu bilayers, we demonstrate that the measured anomalous Hall effect should not follow the scaling laws derived from skew scattering or side-jump mechanism due to the short-circuit and shunting effects of the non-magnetic layers.

*Author to whom correspondence should be addressed. Electronic address: phxxz@ust.hk

Anomalous Hall effect (AHE) has been intensively studied recently for the renewed interest in its origin (or mechanism) in magnetic materials [1-7] and its important role in the characterization of the magnetic state in diluted magnetic semiconductor materials [8-10]. The Hall resistivity in ferromagnetic materials is usually described by the following empirical equation [11]:

$$\rho_{xy} = R_O H + 4\pi R_S M, \qquad (1)$$

where $R_O H$ and $4\pi R_S M$ are ordinary Hall effect and AHE respectively; $R_O$ and $R_S$ are according the ordinary and anomalous Hall coefficient; M is the magnetization of the sample.

The mechanism(s) of AHE was first proposed by Karplus and Luttinger[12] to be intrinsic, which led to a quadratic dependence of AHE resistivity, $\rho_{xy}$ (or $R_S$) on longitudinal resistivity $\rho_{xx}$. Smit argued that AHE should be resulted from the scatterings of electrons at imperfections in a crystal, i.e. extrinsic in nature [13]. Based on the extrinsic spin-orbit interaction, skew-scattering (SS) and side-jump (SJ) mechanisms were proposed by Smit [13] and Berger [14]. Based on the skew scattering and side-jump models, the scaling between $\rho_{xy}$ (or $R_S$) and $\rho_{xx}$ was derived theoretically, i.e.

$$\rho_{xy} \sim \rho_{xx}^n \qquad (\text{or} \qquad R_S \sim \rho_{xx}^n), \qquad (2)$$

where exponent n =1 and 2 correspond to skew scattering and side-jump respectively. Since then, most experimental results have been interpreted based on the value of n obtained by fitting experimental data to eq. 2. The intrinsic AHE resurfaced only a few years ago [1-7].

When both mechanisms contribute, the total AHE may be written as

$$\rho_{xy} = a\rho_{xx} + b\rho_{xx}^2, \qquad (3)$$

where *a* and *b* are constants for a given material. In practice, the experimental data are usually fitted to the power law, $\rho_{xy} \sim \rho_{xx}^n$, and the fitted-value of exponent n is taken as a criteria to judge which mechanism dominates. Very different values of n have been obtained in the different materials. For example, n=1 in low-resistivity dilute alloys [11] and n= 3.7 in heterogeneous Co-Ag films [15].

AHE in the multilayer systems has been intensively studied, because the giant magnetoresistance (GMR) observed in these systems is believed to be closely related to the spin-dependent scattering and the investigation of AHE may provide extra information for understanding GMR [16-26]. However, very different values of n were reported. Khatua *et al.* [18,19] found that n can be surprisingly large (2.8-3.3) in the Fe/Cr multilayers that exhibit GMR, if values of $\rho_{xx}$ at zero magnetic field and values of $\rho_{xy}$ at saturation field are used in the fitting. When values of $\rho_{xy}(H)$ and $\rho_{xx}(H)$ at the same magnetic fields were used in fitting the data to eq. 2, they obtained n=1.96±0.02. They then claimed that side-jump mechanism is dominant in these Fe/Cr multilayers. Song *et al.* [25] found n=2.6 in the Fe/Cr mulatilayers deposited by the electron-beam and claimed that the interface scattering resulted in the large exponent. Interestingly, Tsui *et al.* [26] found that the scaling law works well in the molecular beam epitaxy grown Co/Cu multilayers, i.e. $R_S \sim \rho_{xx}^2(0)$. However, the theoretical study of AHE in multilayer systems by Zhang [23] showed that $\rho_{xy} \propto \rho_{xx}^2$ is valid only when the mean free path of electrons is less than the layer thickness. It is evident that the origin or mechanism of AHE in ferromagnetic/non-magnetic multilayer systems has not been well understood yet and that it is not clear whether the theory developed for homogeneous materials can directly be applied to the heterogeneous materials.

In this letter, we would like to address, in a different angle, the applicability of the theory to the multilayer systems using simple bilayers. In a multilayer sample made of ferromagnetic (FM) and nonmagnetic (NM) layers, the anomalous Hall voltage is only generated in ferromagnetic layers and the nonmagnetic layers actually form conducting path in the transverse direction that releases the electric charges accumulated by AHE in the FM layers. In other words, the NM layers reduce the AHE voltage, which may be called a short-circuit effect (SCE). On the other hand, both the FM and NM layers conduct current, which certainly reduces the total current passing through the FM layers, and consequently reduce the total AHE voltage for a given current. This shunting effect may become dominant when the NM layers are thicker than the FM layers and/or the resistivity of NM layers is smaller that that of FM layer.

Scaling between the AHE and the longitudinal resistivities is based on the fact that both of them arise from scattering of electrons at the imperfections in a FM material [27]. In a multilayer specimen, when the AHE generated in the FM layers is short-circuited by the NM layers, a non-scattering ingredient is then introduced into the whole AHE. Similarly, the measured longitudinal resistivity of the whole sample also contains the contribution from the NM layers. Therefore, both the $\rho_{xy}$ and $\rho_{xx}$ may not really originate from the same mechanism, the scattering of electrons by the imperfections and a simple scaling between these two resistivities will not be a proper criterion to justify the origin of AHE. We will use Fe/Cu bilayers of different layer thicknesses, the simplest model system, to demonstrate the importance of the short-circuit effect and shunting effect in the multilayer systems.

Bilayers of Fe(140 nm)/Cu(x) with 0≤x≤47 nm were fabricated using the magnetron sputtering on the glass, single crystal quartz disk (5 mm in diameter and 0.1mm thick) and Kapton substrates. Films on the Kapton and quartz disks were used for magnetic measurements. The base pressure was always below $2 \times 10^{-7}$ Torr, and the working pressure of Ar gas was usually kept at $3 \times 10^{-3}$ Torr. During the deposition the substrates were at room temperature. Masks were used to fabricate patterned films on the glass substrates for (magneto-) electrical transport measurements ( $\rho_{xy}(H,T)$ and $\rho_{xx}(H,T)$ ). In all the bilayer samples, Fe layers were immediately deposited on the Cu layers after the Cu layer deposition. Therefore, there will not be any oxide interface between Cu and Fe layers. All the transport measurements were carried out on a Quantum Design Physical Property Measurement System (PPMS). After the magnetic saturation, the $\rho_{xy}(H)$ will be dominated by the OHE, the first term in eq.1. In the following we will focus on the AHE resistivity, $\rho_{xyA}(H) = \rho_{xy}(H) - R_O H$. The structure and morphology of the thin Cu films were studied using a transmission electron microscopy (TEM).

Fig.1 shows $\rho_{xyA}(H)$ at 5K for Fe(140 nm)/Cu(x) bilayers with different thickness of Cu layers (only the positive fields data are shown here for clarity). The linear dependence of $\rho_{xyA}(H)$ on H below H=16 kOe can be easily ascribed to the field-forced rotation of the magnetization from in-plane to the field direction, which indicates a good in plane anisotropy originated from the demagnetization effect. The most interesting feature of this set of data is that as the Cu layer becomes thicker, the AHE decreases significantly as expected, except two samples with 1 and 2 nm thick Cu layers. The saturated $\rho_{xyA}(H)$ extracted from fig.1 as a function of thickness of the Cu layers are plotted in the inset of fig.2, in which the data obtained at 180K and

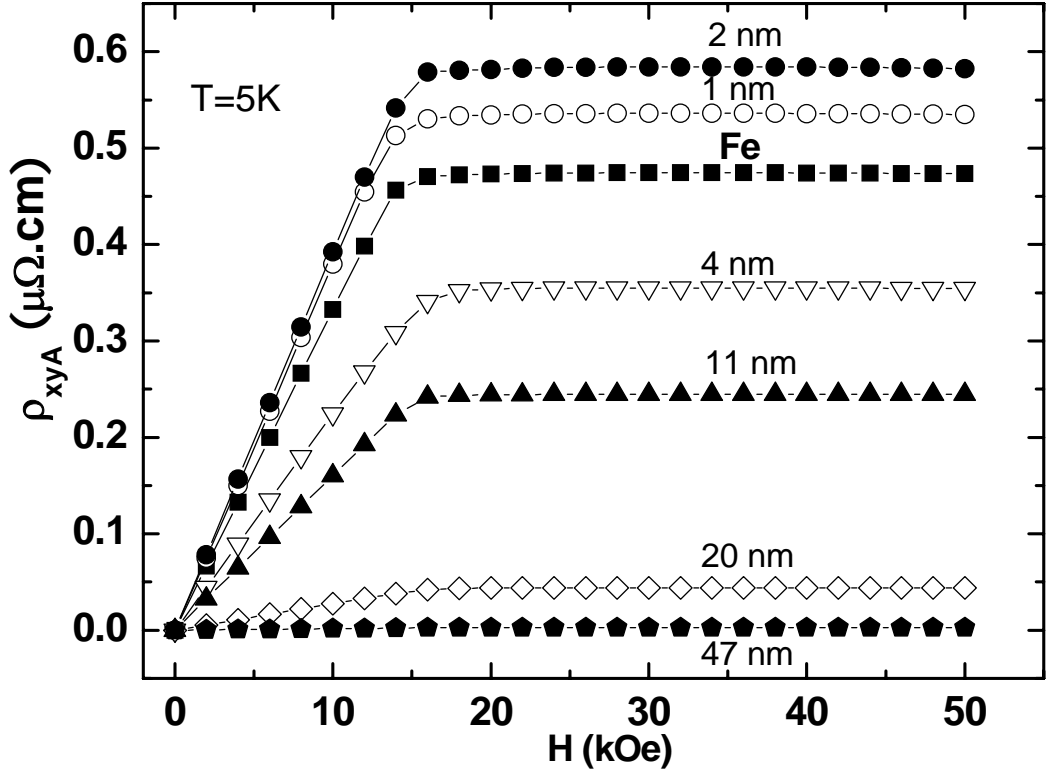

Fig.1. Field dependent anomalous Hall resistivity for Fe/Cu bilayers with different Cu layer thicknesses measured at 5K. The thickness of Fe layer is fixed at 140 nm, the thicknesses of Cu layers change from 0 to 47 nm.

300 K are also plotted. The saturated $\rho_{xyA}(H)$ obtained at different temperatures show an almost identical dependence, indicating the magnetic state of the film depends very weekly on temperature. The overall trend of the curve is that with increasing thickness of the Cu layer, the AHE resistivity decreases. When Cu layer is thicker than 30 nm, AHE becomes very small, e.g. the ratio of saturated $\rho_{xyA}(H)$ for single Fe layer and a bilayers with Cu layer 47 nm thick, reaches 700. This strong reduction of the AHE resistivity indicates the existence of the short-circuit or/and shunting effect due to Cu layer.

To analyze the each contribution to the reduction of AHE, let us assume the Fe/Cu bilayers to be two independent resistors in parallel [28]. In this simple model, the current flow through the Fe layer is $IR_{Cu}/(R_{Cu}+R_{Fe})$, where $I$ is the total current passing through the bilayers. Since the AHE is generally calculated using the total current from the experimental data, the obtained AHE should then be equal to the AHE resistivity of pure Fe film multiplied by a factor $R_{Cu}/(R_{Cu}+R_{Fe})$, i.e.

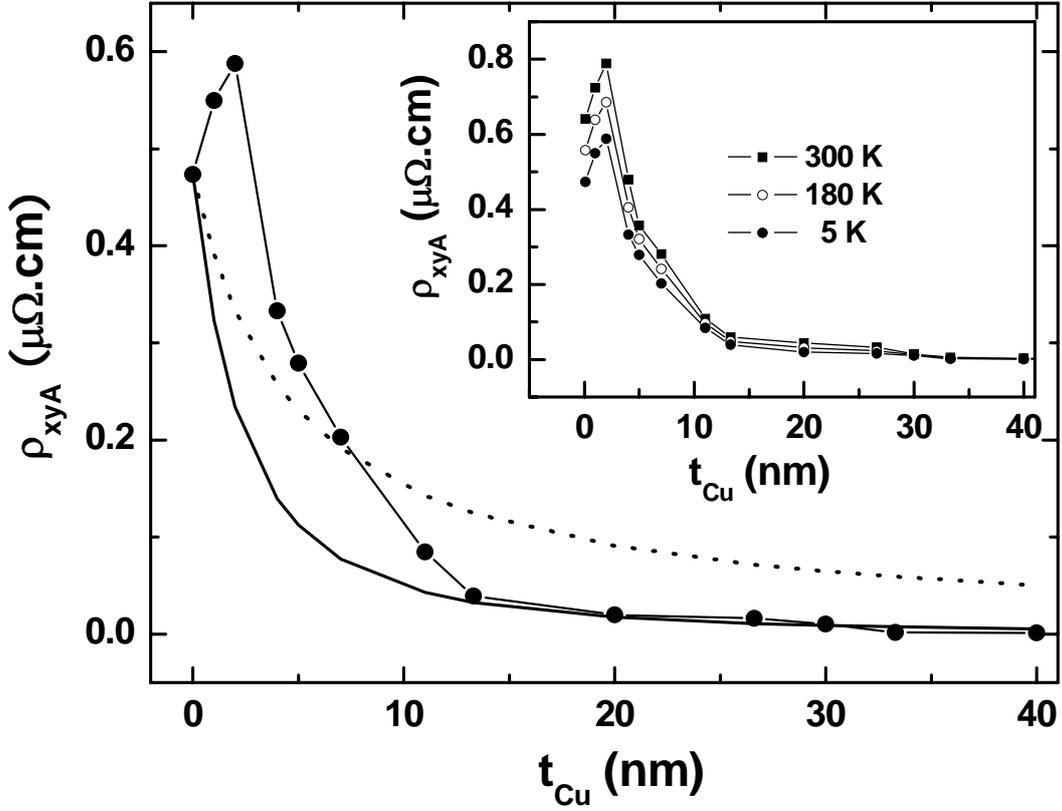

Fig.2. Saturated anomalous Hall resistivity of Fe/Cu bilayers as a function of Cu layer thickness measured at 5K. The dot line and thick solid line are the calculated results by considering only the shunting effect and both the shunting and short-circuit effect respectively. The inset is the saturated anomalous Hall resistivity of Fe/Cu bilayers as a function of Cu layer thickness measured at 5, 180 and 300 K.

$$\rho_{xyA,bilayer} = \rho_{xyA,Fe} \cdot \frac{1}{\frac{\rho_{Fe}}{\rho_{Cu}} \cdot \frac{t_{Cu}}{t_{Fe}} + 1}, \qquad (4)$$

where $\rho_{Fe}$ and $\rho_{Cu}$ are longitudinal resistivities of pure Fe and Cu layers; $t_{Fe}$ and $t_{Cu}$ are their thicknesses. Using eq. 4 and the AHE resistivity of pure Fe film and the experimental data of $\rho_{Fe}/\rho_{Cu}$ for the bilayer samples, we obtained the dependence of saturated $\rho_{xyA}(H)$ on $t_{Cu}$ for $t_{Fe}=140$ nm at T=5 K (doted line in fig. 2). Shown in fig. 2 are the calculated and experimental data of the saturated $\rho_{xyA}(H)$ at 5 K. It is clear that the reduction of AHE can not be ascribed only to the shunting effect.

To understand the short-circuit effect, we will consider that Fe and Cu layers in the bilayers sample are two independent layers. In the transverse direction (perpendicular to the current), a Hall voltage is established due to the AHE in the Fe layer. The Hall voltage is actually the equilibrium of two effects. The first one is a transverse current caused by the skew scattering or/and side jump effect, which accumulates the electrons in one side of the sample. This process does not stop if both the longitudinal current and magnetic field are being applied, even a stable Hall voltage is already established. This picture should be much different from the ordinary Hall effect, where the Lorenz force is balanced by the electric force of Hall electric field, and no transverse current exits. This is because both forces are classical forces and can be balanced. However, in a FM sample, the transverse electric field due to the anomalous Hall voltage will continuously drive the electrons to another side of the sample, which leads to a transverse current in the opposite direction to the current due to skew scattering or side-jump effect. The balance of two transverse currents leads to a stable AHE voltage. A bilayers sample can be considered as a closed circuit of two resistors ($R_{Fe}$, and $R_{Cu}$) connected in series and a total electrical potential of $V_{xy,Fe}$, the anomalous Hall voltage of Fe layer. Therefore, the measured Hall voltage of bilayers can be easily calculated as $V_{xy,Fe} R_{Cu} / (R_{Cu} + R_{Fe})$. It is clear that the measured AHE resistivity has the same dependence on the thickness of Cu layer as described by eq. 4, if only short-circuit effect is considered.

According to the analysis, when both shunting effect and short-circuit effect are considered simultaneously, the measured AHE resistivity will depend on the thickness of Cu layer as

$$\rho_{xyA,bilayer} = \rho_{xyA,Fe} \cdot \frac{1}{\left(\frac{\rho_{Fe}}{\rho_{Cu}} \cdot \frac{t_{Cu}}{t_{Fe}} + 1\right)^2} \quad . \tag{5}$$

The thick solid line in fig. 2 is calculated using eq. 5. It is apparent that the calculated data are in a good agreement with the experimental data for $t_{Cu} \geq 13$ nm. This clearly confirms our analysis that non-magnetic metal layer in the bilayers (or multilayers) will reduce the AHE due to the shunting and short-circuit effects.

In order to directly verify the short-circuit effect without the interference of the shunting effect, we used a variable resistor to substitute the Cu layer connected to the Hall bar of a 150 nm pure Fe layer. The circuit is shown in the inset of fig. 3. The change in the resistance R can be considered as the change in the thickness of the Cu films. When the resistance R is infinite (just disconnecting R), the measured Hall voltage is $V_{AHE,Fe}$, the AHE of pure Fe layer. When the resistor is not infinite, then the measured voltage is given by

$$V_m = \frac{R}{R+R_{Fe}} V_{AHE,Fe} \qquad (6)$$

where $R_{Fe} = 21$ Ω is the resistance of the Fe film between the two Hall contacts. Shown in fig. 3 is the measured AHE resistivity extracted from $V_m$ (symbols) and calculated $V_m$ using eq. 6 (lines) as a function of applied magnetic field with different values of R. The agreement between the calculated and the measured data is surprisingly good. This simple experiment clearly demonstrated the short-circuit effect of the non-magnetic layer in the bilayers (or multilayers systems), because the shunting effect was completely removed.

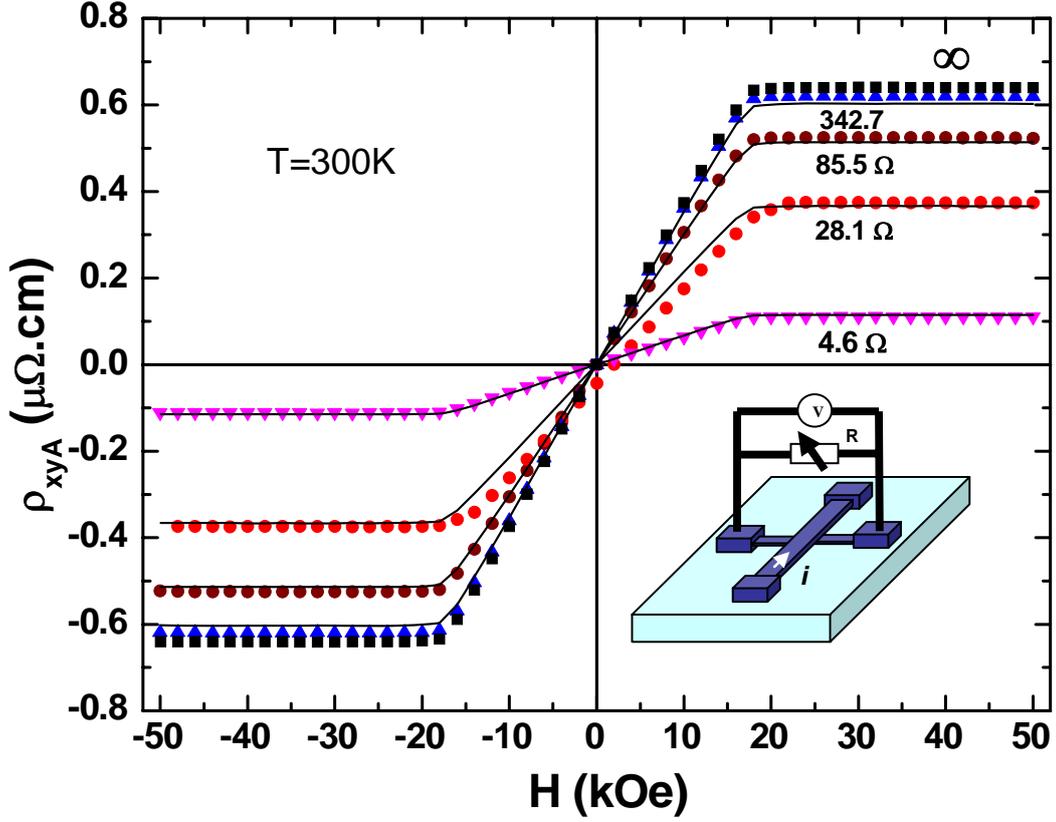

Fig.3. Field dependent anomalous Hall resistivity for Fe film of 150 nm that is connected to a variable resistor R (inset). The lines are the calculated by considering the short-circuit effect due to the variable resistor R.

Now let us go back to fig.2. One should note that a peak in AHE appears at $t_{Cu}$ ~2 nm and the peak value is higher than that for pure Fe layer, which seems contrary to what we discussed above. Actually, if one accepts that scattering of electrons by imperfections can lead to AHE (even partially), the origin (or mechanism) of the peak can be easily understood as follows. It is well known that the film growth at the initial stage with sputtering or evaporation technique is not the same as that with molecular beam epitaxy (MBE) growth where a film grows monolayer by monolayer. At the initial stage of growth of a metal film by sputtering /evaporation deposition, the film is usually not a continuous film. Individual clusters nucleate first, then grow and form percolated network [29,30]. Finally, the holes in the network are filled up. In other words, only when the average thickness is thicker than a threshold, a film becomes continuous. Of course this threshold depends on the material and fabrication conditions (*e.g.* substrate temperature and deposition rate etc.). Therefore, a very thin Cu layer can be considered as individual or agglomerated clusters, which will lead to a very rough surface for the latterly deposited Fe layer. This will increase the AHE due to increased impurities. With increase of the thickness of Cu layer, the enhancement of AHE increases. However, when a percolated network forms with increasing the average thickness of the Cu layer, the shunting and short-circuit effects will begin to work. The competition between the enhancement and reduction of AHE due to the different mechanisms will certainly results in a peak in AHE. The experimental values of AHE for 2 nm< $t_{Cu}$ <13 nm (fig. 2) are higher than the calculated values can be also ascribed to the roughness of the interface between the Cu and Fe layers. When $t_{Cu}$ > 13 nm, the contribution of the interface become relatively weak. To support above explanation, we have simultaneously deposited the Cu films on the Cu grid covered by thin carbon films for the TEM studies. Shown in fig. 4 are the TEM images for Cu film of different thicknesses. It is clearly seen that the 1 nm film is not a real film, but Cu clusters (not percolated yet). The clusters in the 2 nm sample are much larger than those in 1 nm sample and form a network. The image for Cu layer of 5 nm thick clearly shows a continuous film. These TEM images indeed strongly support our interpretation to the peak and other features of fig. 2.

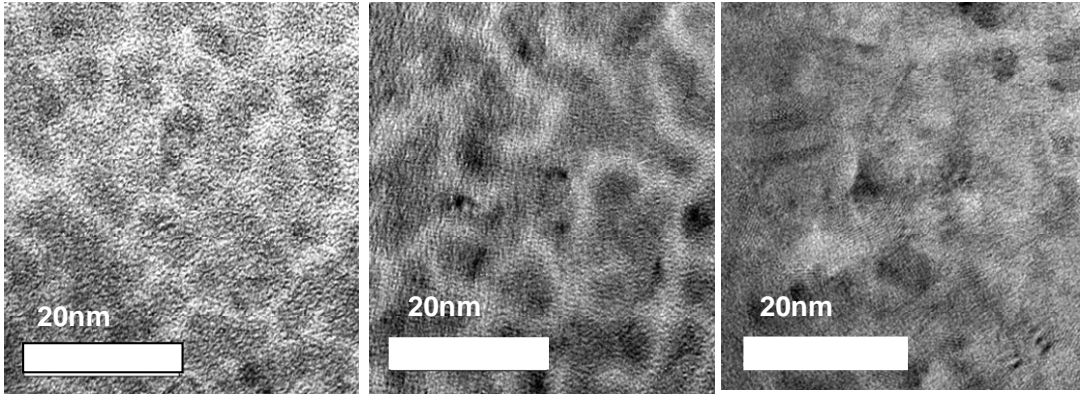

Fig.4. Transmission electron microscopy images of Cu films of different thickness, $t_{Cu}$ ~ 1, 2 and 5 nm.

In summary, we have demonstrated that the non-magnetic layer in the bilayers or multilayers systems can significantly reduce the AHE due to the short-circuit and shunting effects that have a much complicated dependence on the longitudinal resistivity and thickness of both magnetic and non-magnetic layers. Therefore, the scaling laws derived from the skew scattering and side-jump mechanisms are intrinsically not applicable to the multilayer systems.

The work described in this paper was partially supported by grants from the Research Grants Council of the Hong Kong Special Administration Region (Project No. 605704 and 604407) and NSFC (No. 50729101).